\documentclass[prc,twocolumn,showpacs,preprintnumbers,amsmath,amssymb,superscriptaddress,floatfix,nofootinbib]{revtex4}

\usepackage{graphicx}
\usepackage{amsmath}
\usepackage{amsfonts}
\usepackage{amssymb}%
\begin{document}
\title{The $pp \to p \Lambda K^+ $ and $pp \to p \Sigma^0 K^+$ reactions with chiral dynamics}
\author{Ju-Jun Xie} \email{xiejujun@ific.uv.es}
\affiliation{Instituto de F\'\i sica Corpuscular (IFIC), Centro
Mixto CSIC-Universidad de Valencia, Institutos de Investigaci\'on de
Paterna, Aptd. 22085, E-46071 Valencia, Spain}
\affiliation{Department of Physics, Zhengzhou University, Zhengzhou,
Henan 450001, China}
\author{Hua-Xing Chen} \email{hxchen@ific.uv.es}
\affiliation{Instituto de F\'\i sica Corpuscular (IFIC), Centro
Mixto CSIC-Universidad de Valencia, Institutos de Investigaci\'on de
Paterna, Aptd. 22085, E-46071 Valencia, Spain}
\affiliation{Departamento de F\'\i sica Te\'orica, Universidad de
Valencia, Valencia, Spain} \affiliation{Department of Physics and
State Key Laboratory of Nuclear Physics and Technology, \\ Peking
University, Beijing 100871, China}
\author{E. Oset} \email{oset@ific.uv.es}
\affiliation{Instituto de F\'\i sica Corpuscular (IFIC), Centro
Mixto CSIC-Universidad de Valencia, Institutos de Investigaci\'on de
Paterna, Aptd. 22085, E-46071 Valencia, Spain}
\affiliation{Departamento de F\'\i sica Te\'orica, Universidad de
Valencia, Valencia, Spain}

\begin{abstract}
We report on a theoretical study of the $pp \to p \Lambda K^+$ and
$pp \to p \Sigma^0 K^+$ reactions near threshold using a chiral
dynamical approach. The production process is described by
single-pion and single-kaon exchange. The final state interactions
of nucleon-hyperon, $K$-hyperon and $K$-nucleon systems are also
taken into account. We show that our model leads to a fair
description of the experimental data on the total cross section of
the $pp \to p \Lambda K^+$ and $pp \to p \Sigma^0 K^+$ reactions. We
find that the experimental observed strong suppression of $\Sigma^0$
production compared to $\Lambda$ production at the same excess
energy can be explained. However, ignorance of phases between some
amplitudes does not allow to properly account for the
nucleon-hyperon final state interaction for the $pp \to p \Sigma^0
K^+$ reaction. We also demonstrate that the invariant mass
distribution and the Dalitz plot provide direct information about
the $\Lambda$ and $\Sigma^0$ production mechanism, and may be tested
by experiments at COSY or HIRFL-CSR.
\end{abstract}

\maketitle

\section{Introduction} \label{Intro}

The $pp \to p \Lambda K^+$  and $pp \to p \Sigma^0 K^+$ reactions
close to threshold have been advocated as a source of information on
the $p \Lambda$ interaction due to a clear enhancement of the $p
\Lambda$ invariant mass distribution close to threshold
\cite{Siebert:1994jy} with respect to a pure phase space
expectation. The effect of the $p \Lambda$ final state interaction
(FSI) was already studied in
Refs.~\cite{Gasparian:1999jj,Sibirtsev:2005mv}, with a model for the
$pp \to p \Lambda K^+$  and $pp \to p \Sigma^0 K^+$ based on $\pi$
and $K $ exchange and meson baryon amplitudes evaluated with the
Juelich model. Further investigations were carried out in
Ref.~\cite{Hinterberger:2004ra} in terms of the inverse Jost
function and the effective range approximation. More recently the
issue of the FSI in these reactions has been retaken in
Refs.~\cite{Gasparyan:2003cc,Gasparyan:2005fk} using dispersion
relations, and a further experimental research has given more
support to the role of the $\Lambda p$ FSI in these
reactions~\cite{Rozek:2006ct}. A further incursion into the problem
looked for angular distribution as further observable that supported
the importance of the $\Lambda p$ FSI \cite{Sibirtsev:2006uy}. With
suitable parametrizations of the bare amplitude for the $pp \to p
\Lambda K^+$  and $pp \to p \Sigma^0 K^+$ reactions prior to
$\Lambda p$ FSI, a good reproduction of the shapes and ratio of the
cross sections of the two reactions was obtained in a wide range of
energies, considering FSI in the $pp \to p \Lambda K^+$ reaction but
not in the $pp \to p\Sigma^0 K^+$ reaction. A different approach,
with different results on the scattering lengths and effective range
for the $p \Lambda$ interaction is offered in Ref.~\cite{hinterask}.

Within a different approach to the problem, in Ref.~\cite{liuzou}
the authors give an explanation to the $pp \to p K^+ \Lambda$
reaction based on the main mechanism of $N^*(1535)$ excitation
mediated by $\pi, \eta, \rho$ exchange. Previous work on the issue
included contribution from the excitation of the $N^*(1650)$,
$N^*(1710)$ and $N^*(1720)$ \cite{alex,Shyam:2005dw}. In
Ref.~\cite{alex2} the authors show that the consideration of the
final state interaction can make effects similar to the excitation
of the $N^*(1535)$ considered in Ref.~\cite{liuzou}, and the data of
Ref.~\cite{cosi-tof} support the excitation of the $N^*(1650)$
resonance. In a reply to Ref.~\cite{alex2}, the authors of
Refs.~\cite{liuzou2,zoucontri} argue that in the $J/\Psi \to
\bar{p}K^+ \Lambda$ reaction~\cite{yanghx} the $N^*(1535)$ is the
most outstanding signal and they conclude that the inclusion of the
$N^*(1535)$ in the analysis of the $pp \to p\Lambda K^+$ reaction
may reduce the $N^*(1650)$ contribution necessary to reproduce the
data. In our approach, which relies upon pion and kaon exchange and
chiral amplitudes, the $\pi N \to K \Lambda$ amplitude appears in
the scheme, and the unitarization of this amplitude using the chiral
unitary approach produces naturally the  $N^*(1535)$ resonance
\cite{Kaiser:1995cy,Kaiser:1996js,inoue,Nieves:2001wt}, such that we
can make a quantitative statement on its relevance in the $pp \to
p\Lambda K^+$ reaction. On the other hand the $p \Lambda$
interaction close to threshold is very strong
\cite{Dover:1985ba,Maessen:1989sx}, and final state interaction due
to this source is unavoidable in an accurate calculation, and we
also take it into account. We use a dynamical model similar to the
one in Ref.~\cite{Gasparian:1999jj} but we allow all pairs in the
final state to undergo FSI, as a consequence of which we obtain a
contribution from the $N^*(1535)$ using chiral unitary amplitudes.
Our approach also differs from the other approaches on how the FSI
is implemented, and for this we follow the steps of
Ref.~\cite{junkojuan}. In this reference the chiral unitary
approach, where only scattering amplitude are studied (asymptotic
wave functions for $\vec{r} \rightarrow \infty$), is extended to
obtain wave functions for all values of $\vec{r}$ and at the same
time determine form factors and effects of final state interaction
in different reactions.

Furthermore, the experimental total cross section for the $p p \to p
\Sigma^0 K^+$ reaction is strongly suppressed compared to that of
the $p p \to p \Lambda K^+$ reaction at the same excess energy. This
was explained by a destructive interference between $\pi$ and $K$
exchange in the reaction $pp \to p \Sigma^0
K^+$~\cite{Gasparian:1999jj}. In Ref.~\cite{Shyam:2005dw}, the
$\Sigma^0$ strong suppression was reproduced by the inclusion of the
contributions from $N^*(1650)$ resonances in the total cross
sections of both $pp \to p\Lambda K^+$ and $pp \to p\Sigma^0 K^+$
reactions. We also find a reduction of the $pp \to p\Sigma^0 K^+$
cross section relative to that of $pp \to p\Lambda K^+$ at tree
level and show that the nucleon-hyperon FSI can further magnify the
difference.

In next section, we will give the formalism and ingredients for our
calculation. Then numerical results and discussion are given in
section III. Finally, a short summary is given in section IV.

\section{Formalism and ingredients}

\begin{figure}[htbp]
\includegraphics[scale=0.4]{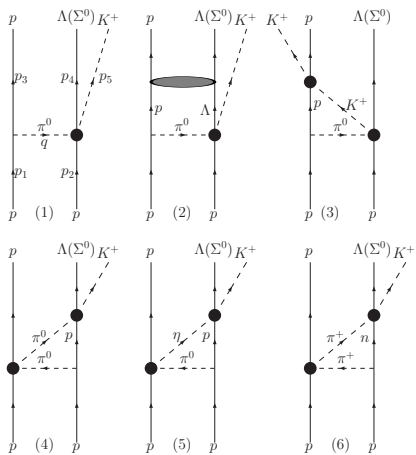}
\caption{The $\pi$ exchange mechanism of the $pp \to p \Lambda
(\Sigma^0) K^+$ reactions. We have also included the final state
interactions. In the first diagram, we show the definitions of the
kinematics ($p_1$, $p_2$, $p_3$, $p_4$, $p_5$, and $q$) that we use
in the present calculation. In addition, we would have the analogous
diagrams permuting the two baryons in the final states.}
\label{Fig:pidiagram}
\end{figure}

\begin{figure}[htbp]
\includegraphics[scale=0.4]{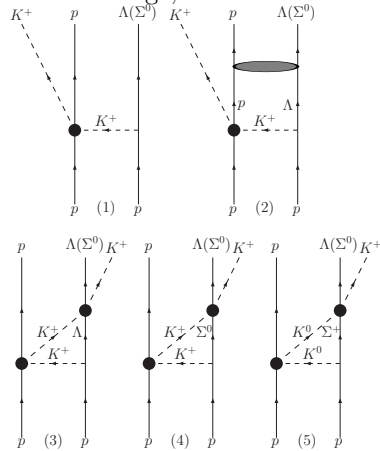}
\caption{The $K$ exchange mechanism of the $pp \to p \Lambda
(\Sigma^0) K^+$ reactions.} \label{Fig:kadiagram}
\end{figure}

To study the reactions $pp \to p \Lambda K^+$ and $p p \to p
\Sigma^0 K^+$, first we investigate the possible reactions
mechanisms in this section. In the reaction at threshold, we
consider the processes involving the exchange of $\pi$ and $K$
mesons as the dominant contributions, as in
Ref.~\cite{Gasparian:1999jj} and other works of the Juelich group.
We show all the possible diagrams exchanging $\pi$ and $K$ mesons in
Fig.~\ref{Fig:pidiagram} and Fig.~\ref{Fig:kadiagram}, respectively.
In the first diagram of Fig.~\ref{Fig:pidiagram}, we show the
definitions of the kinematics ($p_1$, $p_2$, $p_3$, $p_4$, $p_5$,
and $q$) that we use in the present calculation.

The first diagrams of Figs.~\ref{Fig:pidiagram} and
\ref{Fig:kadiagram} show respectively the one $\pi$ and $K$
exchange, without further FSI. The rest of diagrams in
Figs.~\ref{Fig:pidiagram} and \ref{Fig:kadiagram} implement FSI from
meson-baryon and baryon-baryon interactions of the final states.
They are important when we work near the threshold, and should be
taken care of. We note that there are also the corresponding
``mirror'' diagrams where the baryons $p$ and $\Lambda(\Sigma^0)$ in
the final states are permuted with each other. The final states of
the two cases are orthogonal, although they contain the same
particles, hence, there is no interference, but they contribute
equally to the cross sections and this is taken into account.

Assuming an $S$-wave for all the two particle subsystem of the final
states $p \Lambda (\Sigma^0) K^+$, which holds when we work near the
reaction threshold, we can use the conservation of spin and parity
symmetries and obtain that the initial proton-proton system, with
isospin $I=1$, has a total angular momentum $L=1$ and a total spin
$S=1$; therefore, the spin wave functions for the initial
proton-proton system are
\begin{widetext}
\begin{eqnarray}
\hspace{-1.0cm} |pp>= \left\{ \begin{array}{l} |1/2,1/2> |1/2,1/2>,
\vspace{2mm} \cr \frac{1}{\sqrt{2}}(|1/2,1/2>|1/2,-1/2> +
|1/2,-1/2>|1/2,1/2>), \vspace{2mm} \cr |1/2,-1/2>|1/2,-1/2>.
\end{array} \right.
\end{eqnarray}
\end{widetext}

\subsection{Elementary Diagrams}

To write out the total production amplitudes, first we try to write
out the elementary production processes (the first diagrams labeled
by (1) in Figs.~\ref{Fig:pidiagram} and \ref{Fig:kadiagram}). There
are two vertices. One of them is the strong vertex of $\pi NN$ and
$KYN$:
\begin{eqnarray}
f_{\pi NN} \vec \sigma \cdot \vec q {\rm,~~and~~} f_{KYN} \vec
\sigma \cdot \vec q .
\end{eqnarray}
In this process we need several effective interactions for the
strong $\pi NN$ and $KYN$ vertices as shown in the following:
\begin{eqnarray}
f_{\pi^0 pp} &=& \frac{D+F}{2 f_{\pi}}, \\
f_{\pi^+ pn} &=&  \frac{D+F}{\sqrt{2} f_{\pi}}, \\
f_{K^+ p\Lambda} &=&  (-\frac{D+F}{\sqrt{3} f_{K}} +
\frac{D-F}{2\sqrt{3} f_{K}}), \\
f_{K^+ p\Sigma^0} &=& \frac{D-F}{2 f_{K}}, \\
f_{K^0 p\Sigma^+} &=& \frac{D-F}{\sqrt{2} f_{K}},
\end{eqnarray}
In the present work, we use the following parameter values:
$f_{\pi}=93$ MeV, $f_K=1.22~f_{\pi}$~\cite{Gasser:1984gg},
$D=0.795$, and $F=0.465$~\cite{Borasoy:1998pe}.

Another ingredient is the two-body meson-baryon scattering amplitude
such as $T_{\pi^0 p \to K^+ \Lambda}$, etc. We use the chiral
unitary theory to calculate them. The loop functions ($G$) and
$T$-matrices for two-body meson-baryon system are well determined by
a fit to the $S_{11}$ and $S_{31}$ partial wave of $\pi N$
scattering~\cite{inoue} and also the $\bar{K}N$
scattering~\cite{osetKN}. Once these two-body amplitudes are fixed
by the $\pi N$ and $\bar{K} N$ scattering data, we can use them in
the present calculation without introducing any new free parameter.

When we work in the center of mass coordinate of the initial states,
we choose the momenta of the initial protons as $(0, 0, q_z)$ and
$(0, 0, -q_z)$, and we can write the elementary diagrams for $\pi$
and $K$ exchange as (diagrams (1) of Figs.~\ref{Fig:pidiagram} and
\ref{Fig:kadiagram})
\begin{eqnarray}\label{eq:Api1}
{\cal A}^1_{\pi} &=& - F_{\pi NN}(q^2) f_{\pi^0 pp} \sigma_z(1) q_z
\frac{i}{q^2-m^2_{\pi}} T_{\pi^0 p \to K^+ \Lambda}, \\
\label{eq:AK1}{\cal A}^1_{K} &=& F_{KYN}(q^2) f_{K^+ \Lambda p}
\sigma_z(2) q_z \frac{i}{q^2-m^2_{K}} T_{K^+ p \to K^+ p},
\end{eqnarray}
where $m_{\pi}$ and $m_K$ are the masses of $\pi$ and $K$ mesons;
$\sigma_z(1)$ and $\sigma_z(2)$ are the spin Pauli matrices acting
on baryon 1 and baryon 2; $F_{\pi NN}(q^2)$ and $F_{KYN}(q^2)$ are
the form factors for the off-shell $\pi$ and $K$ mesons:
\begin{eqnarray}
F_{\pi NN}(q^2) &=&
\frac{\Lambda^2_{\pi}-m_{\pi}^2}{\Lambda^2_{\pi}-
q^2}, \\
F_{KYN}(q^2) &=& \frac{\Lambda^2_{K}-m_{K}^2}{\Lambda^2_{K}- q^2},
\end{eqnarray}
Here $\Lambda_{\pi}$ and $\Lambda_{K}$ are cutoff parameters where
we take them equivalent in order to minimize the number of free
parameters. This value is usually taken around 1 GeV in our
calculations. We shall do a fine tuning of this value, and obtain
this value after comparing our theoretical results with the
experimental data.

Similarly, we can obtain the ``elementary production amplitudes''
for the other diagrams. By this we mean the remnant of the diagram
omitting the meson-baryon or baryon-baryon FSI. They are

\begin{eqnarray}\label{eq:Aother}
\nonumber {\cal A}^2_{\pi} &=& {\cal A}^3_{\pi} = {\cal A}^1_{\pi} , \\
\nonumber {\cal A}^4_{\pi} &=& F_{\pi NN}(q^2) f_{\pi^0 pp}
\sigma_z(2) q_z \frac{i}{q^2-m^2_{\pi}} T_{\pi^0 p \to \pi^0 p},
\\
\nonumber {\cal A}^5_{\pi} &=& F_{\pi NN}(q^2) f_{\pi^0 pp}
\sigma_z(2) q_z \frac{i}{q^2-m^2_{\pi}} T_{\pi^0 p \to \eta p},
\\
\nonumber {\cal A}^6_{\pi} &=& F_{\pi NN}(q^2) f_{\pi^+ pn}
\sigma_z(2) q_z \frac{i}{q^2-m^2_{\pi}}
T_{\pi^+ p \to \pi^+ p}, \\
{\cal A}^2_{K} &=& {\cal A}^3_{K} = {\cal A}^1_{K} , \\
\nonumber {\cal A}^4_{K} &=& F_{KYN}(q^2) f_{K^+ \Sigma^0 p}
\sigma_z(2) q_z \frac{i}{q^2-m^2_{K}} T_{K^+ p \to K^+ p},
\\
\nonumber {\cal A}^5_{K} &=& F_{KYN}(q^2) f_{K^0 \Sigma^+ p}
\sigma_z(2) q_z \frac{i}{q^2-m^2_{K}} T_{K^0 p \to K^0 p} .
\end{eqnarray}

\subsection{Total Amplitude and Final State Interactions}

The total production amplitude ${\cal M}$ can be written into two
parts:
\begin{eqnarray}
{\cal M} &=& {\cal M}_{\pi} + {\cal M}_{K}, \label{totalm}
\end{eqnarray}
where ${\cal M}_{\pi}$ and ${\cal M}_{K}$ are the amplitudes for the
those diagrams involving $\pi$ and $K$ exchange, respectively. We
have the following formulae:
\begin{eqnarray}
{\cal M}_{\pi} &=& {\cal A}^1_{\pi} +\sum^6_{i=2} {\cal A}^i_{\pi}G_{\pi}^iT_{\pi}^i, \label{mpi} \\
{\cal M}_{K} &=& {\cal A}^1_{K} +\sum^5_{i=2} {\cal
A}^i_{K}G_{K}^iT_K^i, \label{mka}
\end{eqnarray}
where ${\cal A}^i_{\pi/K}$ are the elementary production processes
which we have obtained in Eqs.~(\ref{eq:Api1}), (\ref{eq:AK1}) and
(\ref{eq:Aother}). Together with the free two-body meson-baryon
propagators (such as $G^3_{\pi} = G_{K^+ p}$, etc.), baryon-baryon
propagators (such as $G^2_{\pi} = G_{\Lambda p}$, etc.), and the
final state interactions for meson-baryon cases (such as $T^3_{\pi}
= T_{K^+ p \rightarrow K^+ p}$, etc.) and for baryon-baryon cases
($T^2_{\pi} = T_{\Lambda p \rightarrow \Lambda p}$, etc.), we can
easily write the full total production amplitude ${\cal M}$.

As we have discussed above, the meson-baryon $G$-functions and
$T$-matrices have been calculated in the previous
references~\cite{inoue,osetKN}, and here we only need to deal with
the baryon-baryon ones. Somewhat this is not an easy task from the
theoretical point of view. However, we can obtain them using the
experimental data. For this purpose we follow the strategy described
below.

Following the factorized form of the
$T$-matrix~\cite{osetKN,Oller:2000fj}, for the two-body $\Lambda p$
interaction we use the following type of $\Lambda p \to \Lambda p$
scattering amplitude
\begin{eqnarray}
T_{\Lambda p \to \Lambda p} (\sqrt{s_{p\Lambda}}) =
\frac{1}{V^{-1}-G_{\Lambda p} (\sqrt{s_{p\Lambda}})}, \label{tlpchi}
\end{eqnarray}
where $V$ is the $\Lambda p$ potential and $G_{\Lambda p}$ the loop
function for the $\Lambda p$ system,
\begin{eqnarray}
G_{\Lambda p} (\sqrt{s_{p \Lambda}}) &=& i \int
\frac{d^4q}{(2\pi)^4} \frac{M_{\Lambda}}{E_{\Lambda}(\bf{q})}
\frac{1}{\sqrt{s_{p \Lambda}}-q^0-E_{\Lambda}({\bf q})+i \epsilon} \times \nonumber \\
&&
\frac{M_{p}}{E_{p}(\bf{q})} \frac{1}{q^0-E_{p}({\bf q}) + i\epsilon} \nonumber \\
&=& \int \frac{d^3q}{(2\pi)^3}
\frac{M_{\Lambda}}{E_{\Lambda}(\bf{q})} \frac{M_{p}}{E_{p}(\bf{q})} \times \nonumber \\
&& \frac{1}{\sqrt{s_{p \Lambda}}-E_{\Lambda}({\bf q})-E_{p}({\bf
q})+i\epsilon}, \label{Eq:glp}
\end{eqnarray}
which depends on the invariant mass $\sqrt{s_{p\Lambda}}$ of the $p
\Lambda$ system and a cutoff parameter $\Lambda$.

Both $V$ and $\Lambda$ are determined using the experimental data of
the $\Lambda p \to \Lambda p$ reaction. Here we assume that the
potential $V$ for near threshold $\Lambda p \to \Lambda p$ reaction
is spin and energy independent since we find that this is good
enough to explain the existent experiments. The $\Lambda p \to
\Lambda p$ cross section is then assumed to be
\begin{eqnarray}
\sigma_{\Lambda p \to \Lambda p} = \frac{M^2_pM^2_{\Lambda}}{\pi
s_{p \Lambda}} \Big ( {3\over4}|T^{S=1}_{\Lambda p \to \Lambda p}|^2
+ {1\over4}|T^{S=0}_{\Lambda p \to \Lambda p}|^2 \Big ) ,
\end{eqnarray}
with $T^{S=1} = T^{S=0}$ in our assumption.

\begin{figure}[htbp]
\begin{center}
\includegraphics[scale=0.4]{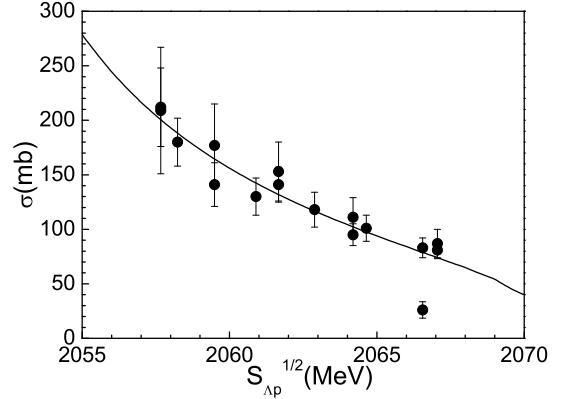}%
\caption{ Total cross sections vs the invariant mass
$\sqrt{s_{p\Lambda}}$ for $\Lambda p \to \Lambda p$ reaction. The
experimental data are taken from Ref.~\cite{lptolpdata}.}
\label{lptolptcs} %
\end{center}
\end{figure}

Then, by comparing the theoretical total cross sections of $\Lambda
p \to \Lambda p$ reaction with experimental data, we can extract the
value of the potential $V$ and the cutoff $\Lambda$ (as shown in
Fig.~\ref{lptolptcs}):
\begin{eqnarray}
V = -6.0 \times 10^{-5} {\rm MeV}^{-2} {\rm,~~and~~} \Lambda=130
{\rm MeV} \, .
\end{eqnarray}

On the other side, the scattering amplitude $T_{\Lambda p \to
\Lambda p} (\sqrt{s_{p\Lambda}})$ can be also expressed by using the
effective range approximation in the field theory as,
\begin{eqnarray}
T_{\Lambda p \to \Lambda p} (\sqrt{s_{p\Lambda}}) = \frac{2\pi
\sqrt{s_{p\Lambda}}}{M_pM_{\Lambda}}
\frac{1}{\frac{1}{\bar{a}}-\frac{1}{2}\bar{r}k^2+ik}, \label{tlpera}
\end{eqnarray}
where $k$ is the momentum of the $\Lambda$ or the $p$ in the
$\Lambda p$ center of mass frame, which is given by
\begin{equation}
k=\frac{\sqrt{(s_{p \Lambda}-(M_{\Lambda}+M_{p})^{2}%
)(s_{p \Lambda}-(M_{\Lambda}-M_{p})^{2})}}{2\sqrt{s_{p \Lambda}}},
\end{equation}

By comparing Eq.~(\ref{tlpera}) with Eq.~(\ref{tlpchi}), we can
easily get
\begin{eqnarray}
\frac{M_{\Lambda}M_p}{2\pi\sqrt{s_{p \Lambda}}}
(\frac{1}{\bar{a}}-\frac{1}{2}\bar{r}k^2) &=& V^{-1}-Re(G_{\Lambda p}(\sqrt{s_{\Lambda p}})), \label{reG} \\
\frac{M_{\Lambda}M_p}{2\pi\sqrt{s_{p \Lambda}}} k &=&-Im(G_{\Lambda
p}(\sqrt{s_{\Lambda p}})), \label{imG}
\end{eqnarray}
with $Re(G_{\Lambda p}(\sqrt{s_{\Lambda p}}))$ and $Im(G_{\Lambda
p}(\sqrt{s_{\Lambda p}}))$ the real and imaginary parts of
$G_{\Lambda p}(\sqrt{s_{\Lambda p}})$, respectively. From
Eq.~(\ref{reG}), we can get the scattering length $\bar{a}=(-1.75
\pm 0.02)$ fm and effective range $\bar{r}=(3.43 \pm 0.07)$ fm by
using the values of $V=-6.0\times 10^{-5}$ MeV$^{-2}$ and
$\Lambda=130$ MeV. Eq.(\ref{imG}) follows exactly since it expresses
unitarity, and the amplitude of Eq.~(\ref{tlpchi}) together with
Eq.~(\ref{Eq:glp}) satisfies unitarity. The value of $\Lambda$ is
relatively small, but this is typical for baryon-baryon interactions
reflecting the long range of one or two-pion exchange. One should
not use this cutoff methods if $\Lambda$ is smaller than the
baryons' momentum in the scattering, but this is not the case in the
range of Fig.~\ref{lptolptcs} or the range needed in the FSI in
Figs. 1(2) and 2(2) for the experiments of
Refs.~\cite{ppdata1,ppdata2}.

\subsection{The Transition between $p p \to p \Lambda K^+$ and $p p \to p \Sigma^0 K^+$}

Without considering the transition between $p \Lambda$ and $p
\Sigma^0$, we can obtain the amplitudes of $p p \to p \Lambda K^+$
and $p p \to p \Sigma^0 K^+$. The results are shown in the following
section where we find that the first one is much larger than the
second one. Therefore, for $p p \to p \Lambda K^+$ we do not need to
consider $p p \to p \Sigma^0 K^+ \to p \Lambda K^+$, but for $p p
\to p \Sigma^0 K^+$ we have to consider $p p \to p \Lambda K^+ \to p
\Sigma^0 K^+$ (We shall be more quantitative below).

After considering the $p \Lambda \to p \Sigma^0$ transition diagrams
(Fig.~\ref{Fig:pidiagram}(2) and Fig.~\ref{Fig:kadiagram}(2)), the
scattering amplitude of $pp \to p\Sigma^0 K^+$ reaction can be
rewritten as two parts with a relative phase $\phi$,
\begin{eqnarray}
{\cal M}={\cal M}_1+e^{i \phi}{\cal M}_2, \label{sigmatm}
\end{eqnarray}
where ${\cal M}_1$ the basic amplitude without including the
transition process, and ${\cal M}_2$ the transition amplitude:
\begin{eqnarray}
{\cal M}_2= {\cal M}_{pp \to p\Lambda K^+}G_{\Lambda p} T_{\Lambda p
\to \Sigma^0 p}. \label{sigmam2}
\end{eqnarray}

\begin{figure}[htbp]
\begin{center}
\includegraphics[scale=0.4]{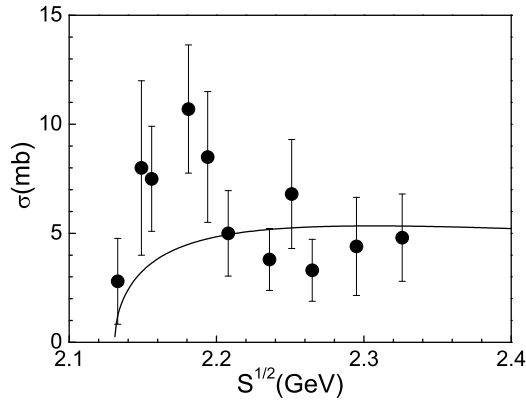}%
\caption{Cross sections of the $\Lambda p \to \Sigma^0 p$ reaction.
The experimental data are taken from Ref.~\cite{sptolpdata}.}
\label{sptolptcs} %
\end{center}
\end{figure}

Here we note that the reaction threshold of $\Sigma^0$ production is
much higher than the $\Lambda$ production, and hence we cannot
calculate ${\cal M}_{pp \to p\Lambda K^+}$ using our previous
result. Again we use the experimental data of the high energy $pp
\to p \Lambda K^+$ reaction~\cite{tofhigh}. Under a pure phase space
expectation we find that a constant value ${\cal M}_{pp \to p\Lambda
K^+} = 1.37 \times 10^{-7}$ MeV$^{-3}$ is consistent with the
experimental data. Similarly, the transition amplitude $T_{\Lambda p
\to \Sigma^0 p}$ is also taken as a constant. Compared with the
experimental data of the $\Lambda p \to \Sigma^0 p$ scattering
\begin{eqnarray}
\sigma_{\Lambda p \to \Sigma^0 p} = \frac{M^2_pM^2_{\Sigma^0}}{\pi
s_{p \Lambda}} \frac{p^{~\text c.m.}_{\Sigma^0}}{p^{~\text
c.m.}_{\Lambda}} |T_{\Lambda p \to \Sigma^0 p}|^2,
\end{eqnarray}
we can get this constant $T_{\Lambda p \to \Sigma^0 p} = 1.53 \times
10^{-5}$ MeV$^{-2}$. We show our theoretical results for the
$\Lambda p \to \Sigma^0 p$ cross section in Fig.~\ref{sptolptcs} by
the solid line. Although the agreement is not perfect, it is
sufficient for the qualitative study that we do in the present study
of the FSI in the $pp \to p\Sigma^0 K^+$ reaction.

It is worth noting that from the total cross section of $pp \to p
\Lambda K^+$ and $\Lambda p \to \Sigma^0 p$ reactions we can only
get the moduli of ${\cal M}_{pp \to p\Lambda K^+}$ and $T_{\Lambda p
\to \Sigma^0 p}$. However, since we need to put a relative phase
$\phi$ between ${\cal M}_1$ and ${\cal M}_2$ in Eq.~(\ref{sigmatm}),
the phase of ${\cal M}_{pp \to p\Lambda K^+}$ and $T_{\Lambda p \to
\Sigma^0 p}$ can be absorbed into the parameter $\phi$, that is
introduced in Eq.~(\ref{sigmatm}). So, in Eq.~(\ref{sigmam2}), we
only need to use the modules for ${\cal M}_{pp \to p\Lambda K^+}$
and $T_{\Lambda p \to \Sigma^0 p}$.

The $G_{\Lambda p}$ function in Eq.~(\ref{sigmam2}) should not be
the same as that for $\Lambda p$ scattering of lower energies since
the $\Lambda p$ energy in this case is higher. We can already see
the first problem that we face to get a quantitative description of
the nucleon-hyperon FSI in this case. Here we shall adopt a
phenomenological approach fitting the relative phase $\phi$ and the
cutoff parameter $\Lambda$ of $G_{\Lambda p}$ trying such as to
reproduce the $p p \to p \Sigma^0 K^+$ cross section, keeping in
mind that $\Lambda$ should be of the same order of magnitude as for
lower energies, but not necessarily equal. As a consequence, we will
not claim a precise prediction of this cross section.

\subsection{Total Cross Section}

From Eqs.~(\ref{totalm}) and (\ref{sigmatm}) we can easily get the
invariant amplitude square $|\cal M|^{\text {2}}$, then the
calculation of the total cross section $\sigma (pp \to p \Lambda
(\Sigma^0) K^+ )$ is straightforward.
\begin{eqnarray}
&& d\sigma (pp\to p \Lambda (\Sigma^0) K^+) =
\frac{1}{3}\frac{M^2_p}{F} \sum_{\text {spins}}|{\cal M}|^2 \times \nonumber \\
&& \frac{M_p d^{3} p_{p}}{E_{p}} \frac{M_{\Lambda(\Sigma^0)} d^{3}
p_{\Lambda(\Sigma^0)}}{E_{\Lambda
(\Sigma^0)}} \times \nonumber \\
&& \frac{d^{3} p_{K^+}}{2 E_{K^+}}  \delta^4
(p_{1}+p_{2}-p_{3}-p_{4}-p_{5}) \label{eqtcs}
\end{eqnarray}
where the flux factor is
\begin{eqnarray}
F=(2 \pi)^5\sqrt{(p_1\cdot p_2)^2-M^4_p}.
\end{eqnarray}
Here we want to discuss a bit about the effect of the spin factor
$\sigma_z$ of Eqs.~(\ref{eq:Api1}), (\ref{eq:AK1}) and
(\ref{eq:Aother}). For $S_{\Lambda p}=1$, the initial and final spin
structures are both symmetric, and so the $\sigma_z$ acting on the
first proton and second proton would lead to the same result. For
$S_{\Lambda p}=0$, the initial spin structure is symmetric and the
final is antisymmetric, and so the $\sigma_z$ acting on the first
proton and second proton would lead to the different result: it
gives an extra minus sign when acting on the second proton.

Now we address the question of the contribution to the $pp \to
p\Lambda K^+$ reaction from the transition of $pp \to p\Sigma^0 K^+
\to p\Lambda K^+$ process. By analogy Eq.~(\ref{sigmam2}) this is
given by
\begin{eqnarray}
{\cal M}'_2= {\cal M}_{pp \to p\Sigma^0 K^+}G_{\Sigma^0 p}
T_{\Sigma^0 p \to \Lambda p} . \label{lambda2}
\end{eqnarray}
Then the relative contribution to the main $pp \to p\Lambda K^+$
term is,
\begin{eqnarray}
R_{\Lambda} = \frac{{\cal M'}_2}{{\cal M'}_1},
\end{eqnarray}
where ${\cal M'}_1$ is the main $pp \to p\Lambda K^+$ amplitude
without considering the $p\Sigma^0 \to p\Lambda$ transition.

Similarly, the relative contribution of ${\cal M}_2$ of
Eq.~(\ref{sigmam2}) to ${\cal M}_1$ of Eq.~(\ref{sigmatm}) for the
$pp \to p\Sigma^0 K^+$ reaction is given by
\begin{eqnarray}
R_{\Sigma} = \frac{{\cal M}_2}{{\cal M}_1}.
\end{eqnarray}
Hence the ratio of the two ratios is
\begin{eqnarray}
\frac{R_{\Lambda}}{R_{\Sigma}} = \frac{{\cal M'}_2}{{\cal
M'}_1}\frac{{\cal M}_1}{{\cal M}_2} = \frac{({\cal M}_{pp \to
p\Sigma^0 K^+})^2}{({\cal M}_{pp \to p\Lambda
K^+})^2}\frac{G_{\Sigma^0 p}}{G_{\Lambda p}}.
\end{eqnarray}

The ratio of the amplitudes squared is of the order of the ratio of
the cross sections of the respective reactions for a same excess
energy and $|G_{\Sigma^0 p}|$ is smaller than $|G_{\Lambda p}|$ at
the threshold of $\Lambda p$ (by a factor of about $3.5$). As a
consequent, the contribution of the transition $pp \to p\Sigma^0 K^+
\to p\Lambda K^+$ relative to the main $pp \to p\Lambda K^+$
amplitude is about $100$ times smaller than the contribution of the
$pp \to p\Lambda K^+ \to p\Sigma^0 K^+$ transition relative to the
main $pp \to p\Sigma^0 K^+$ amplitude and can be neglected.

\section{Numerical results and discussion}

With the formalism and ingredients given above, the total cross
section versus the excess energy ($\varepsilon$) for the $pp \to p
\Lambda K^+$ and $pp \to p \Sigma^0 K^+$ reactions are calculated by
using a Monte Carlo multi-particle phase space integration program.
The results for $\varepsilon$ from $0$ MeV to $14$ MeV are shown in
Figs.~\ref{pltcs} and ~\ref{pstcs} with the cutoff
$\Lambda_{\pi}=\Lambda_K=1300$ MeV, together with the experimental
data~\cite{ppdata1,ppdata2} for comparison.

\subsection{The $pp \to p \Lambda K^+$ Cross Section}

\begin{figure}[htbp]
\begin{center}
\includegraphics[scale=0.4]{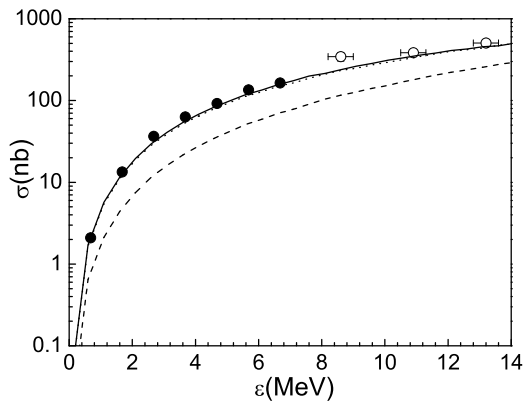}%
\caption{Total cross section vs excess energy $\varepsilon$ for the
$pp \to p \Lambda K^+$ reaction compared with experimental data from
Refs.~\cite{ppdata1} (filled circles) and \cite{ppdata2} (open
circles). Solid and dashed lines show the results from our model
with and without including the $p \Lambda$ FSI, respectively, while
the dotted line represents the results by using the $p\Lambda$ FSI
parameters from Ref.~\cite{hinterask}.} \label{pltcs}
\end{center}
\end{figure}

In Fig.~\ref{pltcs}, we show our results for the case of $pp \to p
\Lambda K^+$ reaction. The solid and dashed lines show the results
from our model with and without including the $p \Lambda$ FSI
(depicted by Fig.~\ref{Fig:pidiagram}(2) and
Fig.~\ref{Fig:kadiagram}(2)), respectively. Alternatively, we also
perform another calculation by using the effective range
approximation for the $T_{\Lambda p \to \Lambda p}$ in the FSI with
the parameters taken from Ref.~\cite{hinterask} ($a_s = -2.43$ fm
and $r_s = 2.21$ fm for the spin of $\Lambda p$ system $S_{\Lambda
p}= 0$; $a_t = -1.56$ fm and $r_t = 3.7$ fm for $S_{\Lambda p} =
1$). The results are shown in Fig.~\ref{pltcs} using the dotted
line. The spin structure of the amplitudes is such that one has a
weight twice bigger for the transition to $S_{\Lambda p} = 1$ than
to $S_{\Lambda p} = 0$. Thus, this fact is implemented by changing
$T_{\Lambda p \rightarrow \Lambda p}$ of Eq.~(\ref{tlpchi}) by that
of Eq.~(\ref{tlpera}) with the $a_i$ and $r_i$ parameters while
using the same $G_{\Lambda p}$ loop function. The weighted cross
section ${1\over3}\big ( 2 \sigma_{S=1} + \sigma_{S=0} \big )$ is
taken.

We can see that in Fig.~\ref{pltcs} both the solid and dotted lines,
which were obtained by including the $p\Lambda$ FSI with different
methods, can reproduce the experimental data quite well for the
excess energy $\varepsilon$ lower than $14$ MeV, but the dashed line
is about two and a half times smaller than the experimental data at
threshold but less than a factor of two smaller than experimental
data at $\epsilon \sim 14$ MeV. This indicates that the $p\Lambda$
FSI is very important in the $pp \to p \Lambda K^+$ reaction close
to threshold. This energy dependence of the FSI is what allows the
determination of the $\Lambda N$ interaction in other approaches
which do not try to get absolute cross
sections~\cite{hinterask,Hinterberger:2005cb}.

It is interesting to note that the spin averaged parameters
$\bar{a}$ and $\bar{r}$ deduced from our approach are very similar
to those of the triplet parameters $a_t$ and $r_t$ in
Ref.~\cite{hinterask} (which have largest weight in the cross
sections),  obtained from the best fit of both the missing mass
spectrum of the reaction $pp \to K^+ +(p \Lambda)$ and the free
$\Lambda p$ scattering by using the standard Jost function approach.
Moreover, our results show that only two parameters in the $p
\Lambda$ interaction are enough to reproduce the current lower
energy experimental data on the $\Lambda p \to \Lambda p$ and $pp
\to p\Lambda K^+$ reactions, but equally good results could be
obtained with the parameters of Ref.~\cite{hinterask}. This
indicates that one should accept the differences between the results
in our approach and those of Ref.~\cite{hinterask} as uncertainties
in the determination of these parameters.

\subsection{The $pp \to p \Sigma^0 K^+$ Cross Section}

For the case of the $p p \rightarrow p \Sigma^0 K^+$ reaction we
need minor changes with respect to the $p p \rightarrow p \Lambda
K^+$ reaction. In the $A_{\pi}^1$ and $A_K^1$ we replace $\Lambda$
by $\Sigma^0$ in the final states. In the other rescattering
diagrams evaluated in Eqs.~(\ref{mpi}) and (\ref{mka}) we replace
$T_\pi^i$ and $T_K^i$ by substituting $\Lambda$ by $\Sigma^0$ in the
final states. This takes into account the $\Lambda p \rightarrow
\Sigma^0 p$ transition, which we argued before was an important term
to consider.

\begin{figure}[htbp]
\begin{center}
\includegraphics[scale=0.4]{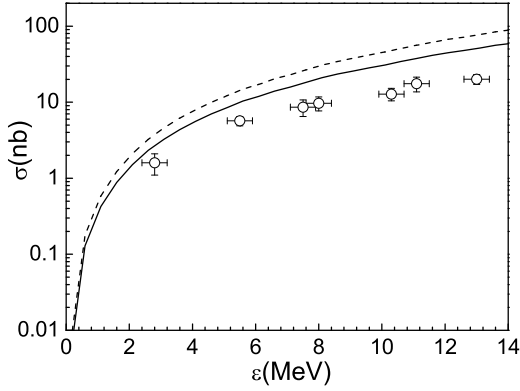}%
\caption{Total cross section vs excess energy $\varepsilon$ for the
$pp \to p \Sigma^0 K^+$ reaction compared with experimental data
from Ref.~\cite{ppdata2}. Solid and dashed lines show the results
from our model with and without including the $p \Lambda \to p
\Sigma^0$ transition diagram, respectively.} \label{pstcs}
\end{center}
\end{figure}

In Fig.~\ref{pstcs}, our results for $pp \to p \Sigma^0 K^+$
reaction are shown. The solid and dashed lines show the results from
our model with and without including the $p \Lambda \to p \Sigma^0$
transition diagrams, respectively. The solid line is obtained using
$\Lambda=310$ MeV and a relative phase $\phi=\pi/2$ between the
basic $pp \to p\Sigma^0 K^+$ amplitude (${\cal M}_1$) and the
transition $pp \to (p\Lambda)K^+ \to (p\Sigma^0)K^+$ amplitude
(${\cal M}_2$). We can see that the FSI mechanism that we have
discussed can indeed induce a reduction of the $pp \to p\Sigma^0
K^+$ reaction, but we mentioned that we do not have control on
$\Lambda$ and $\phi$. On the other hand, we should mention at this
point that the mechanism of $pp \to p \Sigma^0 K^+$ production with
$p \Sigma^0$ FSI, which we have not considered, should be equally
relevant. Indeed, counting simplify cross sections we have at $4$
MeV energy excess $\sigma_{pp \to p \Lambda K^+} \sigma_{p \Lambda
\to p\Sigma^0} \sim 70 \text{mb} \times 5\text{mb}$ while
$\sigma_{pp \to p \Sigma^0 K^+} \sigma_{p\Sigma^0 \to p \Sigma^0}
\sim 7 \text{mb}\times 90 \text{mb}$ which indicate that the
strength of the two amplitudes introduced in the nucleon-hyperon FSI
is similar. The FSI in this case involves two coupled channels in
which we do not have control of interferences. This means that we do
not have at hand enough information within the present formalism to
properly face the nucleon-hyperon FSI in this case. This stated, the
exercise done indicates that this FSI could account for the
difference of our results without this FSI and the data. On the
other hand, we also found that the strong reduction of the $pp \to p
\Sigma^0 K^+$ cross section with respect to the $pp \to p \Lambda
K^+$ one is described by our model at a semiquantitative level.

\subsection{Invariant Mass Spectra and Dalitz Plot}

\begin{figure}[htbp]
\begin{center}
\includegraphics[scale=0.4]%
{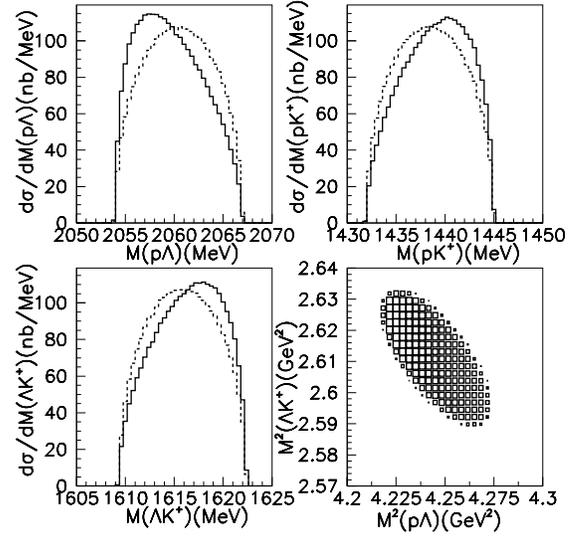}%
\caption{The invariant mass spectra and the Dalitz Plot for the $pp
\to p\Lambda K^+$ at excess energy $\varepsilon =13$ MeV with the
contributions from the full amplitude (solid
curve), compared with pure phase space distributions (dashed curve).}%
\label{imdplambda}%
\end{center}
\end{figure}

\begin{figure}[htbp]
\begin{center}
\includegraphics[scale=0.4]%
{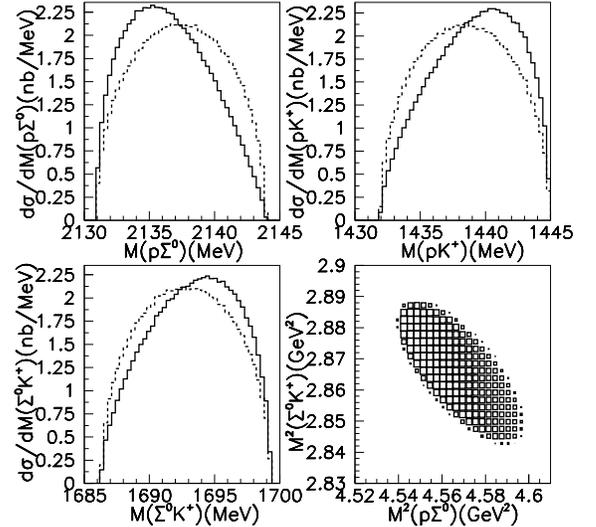}%
\caption{The invariant mass spectra and the Dalitz Plot for the $pp
\to p\Sigma^0 K^+$ at excess energy $\varepsilon =13$ MeV with the
contributions from the full amplitude (solid
curve), compared with pure phase space distributions (dashed curve).}%
\label{imdpsigma}%
\end{center}
\end{figure}

In Figs.~\ref{imdplambda} and \ref{imdpsigma}, we give our
prediction for the invariant mass spectra and the Dalitz Plot for
$pp \to p\Lambda K^+$ and $pp \to p\Sigma^0 K^+$ reactions at excess
energy $\varepsilon =13$ MeV. The dashed line reflects the pure
phase space, while the solid lines include the full amplitudes. The
$p\Lambda$ FSI is very strong and can be seen in the invariant mass
distribution of the $p\Lambda$ system in Fig.~\ref{imdplambda}.

In Fig.~\ref{imdplambda}, the invariant mass distribution of
$K^+\Lambda$ is interesting and counterintuitive. Indeed, we have
introduced explicitly the FSI of the $K^+ \Lambda$ state, which is
dominated by the $N^*(1535)$ below threshold in our approach. We
should expect that the solid line, accounting for FSI should be
shifted to lower invariant masses as a consequence of the presence
of the $N^*(1535)$ below threshold. However the opposite effect is
observed. This is a consequence of the strong effect of the $\Lambda
p$ FSI and we have observed that removing this $\Lambda p$ FSI the
solid line in the $K^+ \Lambda$ mass distribution is indeed shifted
to lower invariant mass which respect to phase space.

In Fig.~\ref{imdpsigma} we have taken the nucleon-hyperon FSI from
the $p\Lambda \to p \Sigma^0$ transition discussed in the text. We
can see features from FSI in the invariant mass distribution that
can help understand these effects better by comparison with future
experiments.

The invariant mass spectra and the Dalitz Plot in
Figs.~\ref{imdplambda} and \ref{imdpsigma} are the direct
information about the $\Lambda$ and $\Sigma^0$ production mechanism
and may be tested by experiments at COSY or HIRFL-CSR.

\section{Summary}

We have made a theoretical study of the $pp\rightarrow p\Lambda K^+$
and $pp\rightarrow p\Sigma^0 K^+$ reactions involving $\pi$ and $K$
exchange and implementing final state interactions of any of the two
hadron pairs in the final states. The amplitudes and loop functions
involved have been obtained using the chiral unitary approach for
meson-baryon interactions~\cite{inoue,osetKN}. The aims were two.
First we wanted to see that the theory provides a fair description
of the cross sections for these two reactions, including the factor
around 30 smaller cross section of the $pp \rightarrow p \Sigma^0
K^+$ reaction than for the $pp \rightarrow p \Lambda K^+$ one at
similar excess energies. We found that the theory indeed
accomplished that qualitatively.

On the second hand, we wanted to see the effect of the final state
interaction, and eventually determine the $\Lambda p \rightarrow
\Lambda p$ low energy parameters, scattering length and effective
range. Here we also succeeded and found reasonable parameters
compatible with the low energy $\Lambda p \rightarrow \Lambda p$
transition cross section and the $pp\rightarrow p \Lambda K^+$ cross
section. These results are also compatible with those determined in
a recent empirical analysis~\cite{hinterask}, but we were able to
show that there are intrinsic uncertainties in the determination of
these parameters from these data, in particular the separation of
the results for the $S=1$ and $S=0$ $\Lambda p$ systems. For the
case of the $p p \rightarrow p \Sigma^0 K^+$ reaction, with much
smaller cross section than the $pp\rightarrow p \Lambda K^+$
reaction, we could obtain qualitative results, but the final state
interaction was influenced by the $\Lambda p \rightarrow \Sigma^0 p$
transition which required extra information than the one deduced and
used in the $pp\rightarrow p \Lambda K^+$ reaction close to
threshold. In this case we have implemented the FSI by introducing a
couple of parameters (a phase and a cutoff) to the data, within a
reasonable range. A better agreement with experiment could be found,
but certainly one does not obtain a quantitative theoretical
prediction. We also show that one should also consider the
nucleon-hyperon FSI from the $p \Sigma^0 \to p\Sigma^0$ amplitude,
but one would not know the interference between the two mechanisms.
The results obtained for $a$ and $r$ for the $\Lambda N$ interaction
at low energies are valuable as empirical determinations of these
data, in line with other determinations. On the other hand, by using
realistic amplitudes extracted from the chiral unitary approach, we
could also show that a determination of the absolute value of the
cross sections is possible, in line of similar findings with
amplitudes from the Juelich model.

We also made predictions for the invariant mass distributions and
Dalitz plots that can be used for comparison with future
experiments.

\section*{Acknowledgments}

This work is partly supported by DGICYT contracts No. FIS2006-03438,
FPA2007-62777, the Generalitat Valenciana in the program PROMETEO
and the EU Integrated Infrastructure Initiative Hadron Physics
Project under Grant Agreement No. 227431. Ju-Jun Xie acknowledges
Ministerio de Educaci\'{o}n Grant SAB2009-0116.

\end{document}